\documentclass[sigconf,nonacm]{acmart}

\AtBeginDocument{%
  }

\usepackage{colortbl}
\usepackage{xcolor}
\usepackage{enumitem}
\usepackage{tikz}
\usepackage{url}
\usepackage{hyperref}
\usepackage{booktabs}
\usepackage{array}
\usetikzlibrary{arrows.meta,positioning,fit,calc}

\begin{document}

\title{Retail Product Search: A Practical Approach at Target}

\author{Darshan Sonagara}
\authornote{These authors contributed equally to this work.}
\affiliation{%
  \institution{Data Sciences, Target Corporation}
  \city{Brooklyn Park}
  \state{MN}
  \country{USA}}
\email{darshan.sonagara@target.com}

\author{Qujiaheng Zhang}
\authornotemark[1]
\affiliation{%
  \institution{Data Sciences, Target Corporation}
  \city{Brooklyn Park}
  \state{MN}
  \country{USA}}
\email{qujiaheng.zhang@target.com}

\author{Ankit Singh}
\authornotemark[1]
\affiliation{%
  \institution{Data Sciences, Target Corporation}
  \city{Sunnyvale}
  \state{CA}
  \country{USA}}
\email{ankit.singh2@target.com}

\author{Alex Li}
\affiliation{%
  \institution{Data Sciences, Target Corporation}
  \city{Sunnyvale}
  \state{CA}
  \country{USA}}
\email{alex.li@target.com}

\renewcommand{\shortauthors}{Sonagara et al.}

\begin{abstract}
Search is one of the most important features in e-commerce, directly driving customer engagement and business growth. A good product search system must show both relevant and desirable results. However, retail search presents unique challenges. User intent can range from exact matches to open-ended discovery. Search systems must also balance multiple goals, such as relevance, revenue, and profit, while keeping response times low. Traditional keyword-based methods often fall short in handling natural language or semantic queries. Vector search helps alleviate these issues, but it can miss key intent signals or return low-precision results. In this paper, we present the design of a hybrid search system at Target that combines lexical and vector search. We describe our approach to data processing, embedding training, precision control for the final result set, multi-channel result fusion (where we compared fusion strategies and adopted weighted interleaving), and the performance optimizations used to maintain low latency for production deployment. Our method improves offline evaluation metrics, and in online A/B testing it raised click-through rate by 0.97\%, order conversion by 0.98\%, and demand per visitor by 1.10\% over lexical-only search, while roughly halving zero-result searches. The resulting system is deployed at scale and serves millions of guests daily.
\end{abstract}



\maketitle

\section{Introduction}
\label{sec:introduction}

Search is a key driver of revenue and user satisfaction in e-commerce. Users who search usually arrive with a specific need, and retrieval quality strongly influences whether they engage with the results and complete a purchase \cite{nigam2019semanticproductsearch,ai2017hierarchical}. When they find the right item quickly, they convert at a much higher rate than casual browsers, which makes an effective search system central to both user experience and business outcomes.

Retail search has its own difficulties. Queries range from precise requests such as ``organic milk 1/2 gallon'' to open-ended ones such as ``cocktail party supplies.'' Different product categories encourage different behavior, from quick repeat purchases to long browsing sessions. Search also has to balance goals that often conflict: showing relevant products, promoting sponsored items, maximizing revenue, and keeping latency low. At large scale, even small increases in latency can hurt system quality and user outcomes \cite{dean2013tail}. A strong system has to understand what the user wants, retrieve the right set of products, and rank the best of them at the top.

Industrial product search usually follows a multi-stage pipeline of query understanding, candidate retrieval, and one or more ranking stages \cite{Manning2008}. Retrieval returns a broad set of plausible candidates, and later stages apply heavier models to reorder them. This split lets fast methods run first and more expensive models run only on a small candidate set. In this paper we focus on the retrieval stage and describe how we combined lexical and vector retrieval into a hybrid system that is both effective and fast.

Lexical retrieval over an inverted index works well when the query and the product text share words. It struggles when they do not, for example when a user's phrasing differs from catalog wording \cite{Furnas1987}. Dense vector retrieval addresses this gap by matching on semantic similarity instead of exact terms \cite{Karpukhin2020,reimers2019sentence}. Dense retrieval has its own problems. It can return loosely related items and miss exact matches that a keyword search would find, and it can add latency if it is not implemented carefully \cite{guo2020accelerating}. How best to combine the two depends on the business context and on system design, and organizations use vector search in different ways. Some limit it to long-tail or vague queries, while others try to replace lexical search altogether \cite{elasticsearch2024}. We run both channels on every query and merge their results, and we show that this hybrid design improves engagement and conversion in online tests \cite{luan2021sparse,Bruch_2023}.

In this paper, we make the following contributions to retail product search:
\begin{enumerate}
    \item A hybrid retrieval architecture that runs lexical retrieval and a fine-tuned dual encoder in parallel, deployed on Target's web and mobile search and serving millions of queries per day.
    \item A labeling scheme that turns engagement logs into positive query-product pairs and two kinds of hard negatives for contrastive training.
    \item A precision control module that filters vector candidates using attributes stated in or predicted from the query.
    \item A comparison of reciprocal rank fusion and weighted interleaving for merging two result lists with little overlap, in which weighted interleaving performed better in online tests.
    \item Offline ablations and 4-week online A/B tests that measure the effect of each component on ranking quality, engagement, and conversion.
\end{enumerate}

The rest of the paper is organized as follows. Section~\ref{sec:literature} reviews related work. Section~\ref{sec:architecture} describes the system architecture and Section~\ref{sec:data} the training data and labels. Section~\ref{sec:semanticModel} covers the semantic model, Section~\ref{sec:precisionControl} precision control, and Section~\ref{sec:mergingStrategy} the merging strategy. Section~\ref{sec:implementationConsiderations} discusses deployment. Section~\ref{sec:experimentsAndResults} reports offline and online results, Section~\ref{sec:conclusion} concludes, and Section~\ref{sec:futureWork} outlines future work.

\section{Literature Review}
\label{sec:literature}

\subsection{Lexical Retrieval}

Early information retrieval relied on Boolean models and exact keyword matching \cite{Manning2008}. These models were efficient but suffered from vocabulary mismatch, where users and catalog descriptions use different words for the same concept \cite{Furnas1987}. Term weighting schemes such as TF-IDF \cite{Salton1988} and BM25 \cite{Robertson2009} improved ranking by accounting for term frequency and inverse document frequency, but they still require the query and the document to share terms.

Query expansion \cite{Voorhees1994} and relevance feedback \cite{Rocchio1971} were proposed to reduce this dependence. Both tend to trade precision for recall and can add noise. That trade-off is costly in e-commerce, where users have little tolerance for irrelevant results and queries range from exact product requests to open-ended browsing.

\subsection{Semantic Search}

Latent semantic indexing \cite{Deerwester1990} was an early attempt to model relationships between terms and documents in a lower-dimensional space, but it was expensive to apply at large scale. Neural dense retrieval \cite{Karpukhin2020} has largely removed these scaling limits, and sentence embedding models have been shown to handle vocabulary mismatch between queries and documents well \cite{reimers2019sentence}.

BERT \cite{Devlin2019} and its variants are widely used to produce contextual embeddings, and in e-commerce, BERT-style models fine-tuned on catalogs and search logs have shown good results. Much of the dense retrieval literature, however, evaluates on general relevance benchmarks rather than on the engagement and conversion metrics that matter to a retailer. Embeddings trained on general text also tend to miss product-specific relationships and purchase intent. Our work targets this gap by training on labels derived from user engagement.

\subsection{Embedding-Based Retrieval in Industry}

Several companies have reported production systems that add embedding-based retrieval to an existing inverted index. Huang et al.~\cite{huang2020embedding} describe unified embeddings for Facebook Search that take the searcher's context into account and are served within a search system built on an inverted index. Zhang et al.~\cite{zhang2020towards} present DPSR at JD.com, a retrieval system that targets both semantic relevance and personalization. Li et al.~\cite{li2021embedding} study the embedding-based retrieval system in Taobao Search and propose methods to improve its relevance and to align training with inference. Magnani et al.~\cite{magnani2022semantic} describe a hybrid system at Walmart that combines an inverted index with neural retrieval to better serve tail queries, which is the setting closest to ours. Our system follows the same broad pattern of running lexical and dense retrieval side by side. It differs in how training labels are built from engagement data, in the precision control applied to the vector channel, and in how the two result lists are merged.

\subsection{Hybrid Search Systems}

Lexical and semantic retrieval have complementary strengths. Lexical search is precise on exact matches and named entities such as brands, while semantic search handles paraphrase and intent. Luan et al.~\cite{luan2021sparse} compare sparse, dense, and attentional representations and show that combining sparse and dense retrieval can outperform either one alone. Hybrid systems differ mainly in how they combine the channels: fusing the results of parallel retrievers, filtering one channel's results with the other, or feeding both kinds of signals to a learning-to-rank (LTR) model.

Score fusion and reciprocal rank fusion (RRF) \cite{Cormack2009} merge ranked lists without additional training. Bruch et al.~\cite{Bruch_2023} analyze fusion functions for hybrid retrieval and find that a convex combination of normalized scores tends to outperform RRF. Score fusion, however, requires scores from different retrievers to be put on a comparable scale, which is hard when BM25 scores and cosine similarities behave very differently. RRF avoids calibration by working with ranks, but it favors items that appear in several lists, and that signal is weak when the lists share few items. LTR models can use both kinds of signals more flexibly, but they usually run as a re-ranking stage after retrieval and add latency. Unified representations such as SPLADE \cite{Formal2021} and COIL \cite{gao2021coil} bring semantic matching into an inverted index. They avoid a separate vector index, but they typically need a transformer pass at query time and larger or specialized index structures, which raises serving cost under tight latency budgets and large catalogs.

\subsection{Challenges in E-Commerce Search}

E-commerce adds requirements beyond general text retrieval \cite{nigam2019semanticproductsearch,ai2017hierarchical}. Product catalogs are heterogeneous: each category has its own attributes, and the importance of each attribute varies, unlike the more uniform document collections common in academic benchmarks. Intent ranges from a specific product to open-ended discovery, and a retrieval system has to serve both. Business constraints such as inventory, profitability, and promotions also shape what should be shown, and they are rarely modeled in research settings.

Our work contributes a hybrid retrieval system that (1) runs in production under strict latency limits, (2) trains embeddings on engagement-derived labels, (3) filters vector candidates with a dedicated precision control module, and (4) merges the two channels with a strategy suited to lists that overlap little.

\section{Architecture}
\label{sec:architecture}

\begin{figure}[t]
    \centering
    \includegraphics[width=\columnwidth]{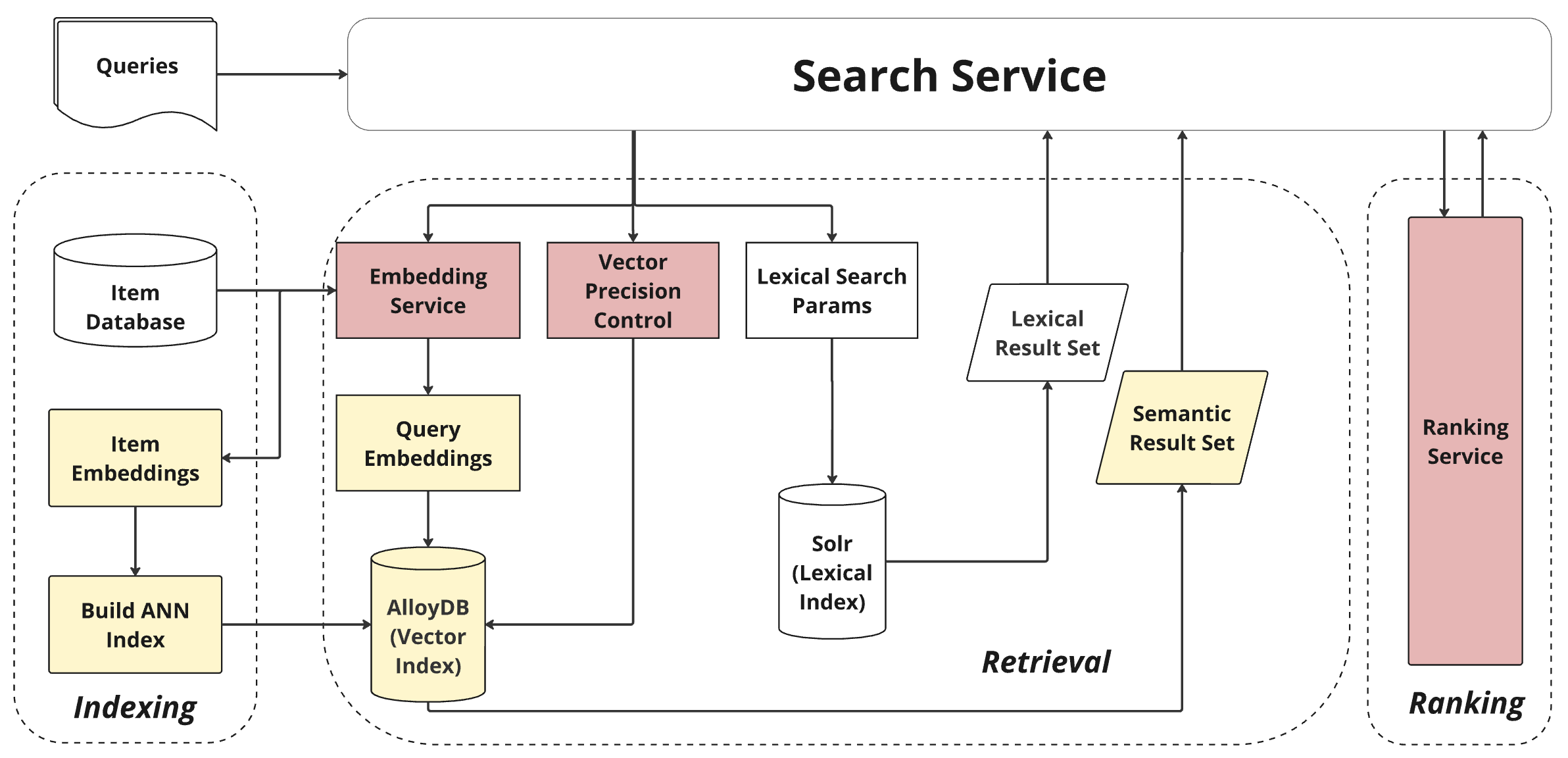}
    \caption{Hybrid search architecture. The indexing pipeline (left) embeds catalog items and builds the vector index. At query time, lexical and vector retrieval run in parallel, and precision control filters the vector results (center). Both result sets then go to ranking (right).}
    \Description{Block diagram. Queries enter a search service, which calls three components in parallel: the embedding service, vector precision control, and lexical search parameters. The query embedding and precision filters are applied to an AlloyDB vector index to produce the semantic result set. The lexical parameters query a Solr index to produce the lexical result set. Both result sets return to the search service, which calls the ranking service. On the left, an indexing pipeline embeds items from the item database and builds the ANN index in AlloyDB.}
    \label{fig:architecture}
\end{figure}

Our hybrid search system combines lexical and semantic retrieval to serve a wide range of shopping intents. Figure~\ref{fig:architecture} shows the main components. Query serving and catalog indexing run as separate pipelines, so the vector index stays current while queries are served.

When a user submits a query, a central search orchestration layer interprets it and builds retrieval requests for the lexical and vector channels. Lexical retrieval runs on a Solr inverted index and gives high precision on token matches and structured attributes. Its configuration supports spell correction, keyword rewriting, facet filtering, and custom boosts driven by query classifiers and fulfillment filters.

In parallel, a dedicated embedding service converts the same query into a dense vector. The service produces both query and item embeddings and runs on inference infrastructure with model versioning and adaptive batching. The query embedding is used to retrieve similar items from an approximate nearest neighbor (ANN) index hosted on AlloyDB, which uses ScaNN as its ANN algorithm \cite{guo2020accelerating}. The index covers millions of product vectors and can match items whose text differs from the query through synonyms, abbreviations, or misspellings.

\begin{sloppypar}A precision control module for vector search removes off-topic vector matches before they reach the result pool. It applies dynamic thresholds, category constraints, and business-specific filters (Section~\ref{sec:precisionControl}).\end{sloppypar}

The results from Solr and AlloyDB are then merged in the ranking layer using weighted interleaving, with channel weights tuned through A/B tests (Section~\ref{sec:mergingStrategy}).

Separately from query serving, an indexing pipeline keeps the vector index up to date. When products are added or updated in the catalog, their attributes are sent to the embedding service, and the resulting vectors are written to the AlloyDB index. The pipeline supports both streaming and batch updates.

Section~\ref{sec:semanticModel} describes the semantic model, Section~\ref{sec:precisionControl} precision control, Section~\ref{sec:mergingStrategy} the merging strategy, and Section~\ref{sec:implementationConsiderations} implementation details.

\section{Data}
\label{sec:data}

The fine-tuning data needs to reflect what users actually choose across many queries, product categories, and platforms. We built a large dataset of labeled query-product pairs from interaction logs aggregated across web and mobile. The logs were anonymized and aggregated at the query-product level, and no personally identifiable information was used.

To keep the training set representative, we applied stratified sampling along two dimensions.
\begin{itemize}
    \item \textbf{Category taxonomy:} We required coverage of all top-level categories and most subcategories. This reduced the share of dominant verticals such as Home and Apparel and increased the share of smaller ones such as Gifts and Electronics.
    \item \textbf{Query frequency:} We grouped queries into head, torso, and tail bands by normalized frequency and gave each band proportional representation, so that tail queries, which have sparse engagement, are not crowded out by head queries.
\end{itemize}

After filtering, de-duplication, and quality checks, the training corpus contained about 20 million labeled query-product pairs. This corpus is the basis for all fine-tuning stages and contributed substantially to the model's performance on torso and tail queries.

\subsection{Labeling}
We built training labels for query-product pairs from historical interaction logs. Click-through rate (CTR), add-to-cart rate (ATC), and conversion rate were aggregated weekly and converted into training targets. Supervision has two parts: positive labels and negative labels.

\subsubsection{Positive Label Construction}
To capture how desirable a product is for a query, we assign each $(\textit{query}, \textit{product}, \textit{week})$ tuple a continuous engagement score that combines the three signals:

\begin{equation}
    S = \alpha \cdot \text{Conversion} + \beta \cdot \text{ATC} + \gamma \cdot \text{CTR}
    \label{eq:formula}
\end{equation}

where $\alpha$, $\beta$, and $\gamma$ are proprietary weights with $\alpha + \beta + \gamma = 1$ and $\alpha > \beta > \gamma > 0$. Their values are withheld for confidentiality. The ordering gives more weight to signals that indicate stronger purchase intent, so conversions count more than add-to-carts, and add-to-carts more than clicks. To avoid fitting to short-term fluctuations, we aggregate over time and keep only query-product pairs whose scores are consistent across multiple weeks. A pair is labeled positive when its score stays above a threshold $T_2$ throughout this window. The value of $T_2$ is withheld for confidentiality. Because of the weighting, a pair with clicks alone needs a much higher click rate to clear $T_2$ than a pair that also leads to conversions.

\subsubsection{Negative Label Construction}
For negatives we combine in-batch sampling with hard negative mining \cite{xiong2021approximate}. With in-batch negatives, each query's positive product is contrasted during training against the positive products of the other queries in the same batch.

We also add hard negatives mined in two ways:
\begin{itemize}
    \item \textbf{Exposure-based hard negatives:} Products with at least $T_1$ impressions for the query (the threshold is withheld for confidentiality) and no engagement of any kind, meaning CTR, ATC, and conversion are all zero. These products were shown often and consistently ignored, which makes them strong negative signals.
    \item \textbf{Similarity-based hard negatives:} Products that often appear among the top-$K$ retrieved candidates for a query but are not positives. They are lexically or semantically close to the query, with similar titles, categories, or attributes, yet do not match what the user wants. They push the model to learn finer distinctions.
\end{itemize}

The two sources cover different failure modes. The first catches products that were shown to users but not wanted, and the second catches products that sit near the query in embedding space but do not satisfy it. Table~\ref{tab:data-ablation} in Section~\ref{sec:experimentsAndResults} shows the effect of adding them.

\section{Semantic Model}
\label{sec:semanticModel}

\subsection{Models}
\subsubsection{Pretrained Model Evaluation}
We first compared several off-the-shelf pretrained transformer encoders to choose a starting point for fine-tuning. Our catalog spans more than 2,500 categories and 7,600 item types, so the model has to generalize broadly while staying accurate. We weighed retrieval quality against inference cost, which Section~\ref{sec:implementationConsiderations} discusses.

For this comparison we built a human-labeled benchmark of 5,000 e-commerce queries, each paired with 200 candidate products graded for relevance by annotators. We use this benchmark only for model selection, because it measures topical relevance independently of engagement data. The fine-tuning experiments in Section~\ref{sec:experimentsAndResults} use a separate engagement-labeled golden set, so the numbers in Table~\ref{tab:pretrained-models} are not comparable with those in Tables~\ref{tab:data-ablation} and~\ref{tab:model-ablation}. Each product was represented by concatenating its \texttt{title}, \texttt{brand}, \texttt{item\_type}, and \texttt{category} into one input string. We encoded queries and products with each model, ranked the candidates by cosine similarity, and measured Normalized Discounted Cumulative Gain (NDCG), which accounts for both relevance grade and rank position.

Table~\ref{tab:pretrained-models} reports the average NDCG for each encoder. \texttt{e5-large-v2} scored highest, followed closely by \texttt{e5-base-v2} and \texttt{e5-small-v2}. The gain from the largest model was small (0.0068 NDCG over \texttt{e5-small-v2}) and came at a much higher inference cost. We therefore chose \texttt{e5-small-v2} \cite{wang2022e5} as the base encoder for fine-tuning, since it gives strong retrieval quality at much lower latency (Section~\ref{sec:implementationConsiderations}).

\begin{table}[!ht]
\centering
\caption{Zero-shot NDCG of pretrained encoders on the human-labeled benchmark (5,000 queries, 200 candidates each), averaged over queries. We selected \texttt{e5-small-v2} for fine-tuning.}
\label{tab:pretrained-models}
\setlength{\tabcolsep}{6pt}
\small
\begin{tabular}{>{\raggedright\arraybackslash}p{0.65\linewidth}
                >{\raggedleft\arraybackslash}p{0.25\linewidth}}
\toprule
\textbf{Model} & \textbf{Average NDCG} \\
\midrule
\texttt{bert-base} & 0.6298 \\
\texttt{msmarco-distilbert-cos-v5} & 0.7510 \\
\texttt{msmarco-bert-base-dot-v5} & 0.7626 \\
\texttt{all-MiniLM-L6-v2} & 0.7703 \\
\texttt{all-mpnet-base-v2} & 0.7707 \\
\texttt{e5-small-v2} & 0.7757 \\
\texttt{e5-base-v2} & 0.7777 \\
\texttt{e5-large-v2} & \textbf{0.7825} \\
\bottomrule
\end{tabular}
\end{table}

\subsubsection{Model Architecture}
Our retrieval model is a Siamese dual encoder, a standard design for semantic search. The query and the product text pass through the same encoder, which shares its weights across both inputs and is initialized from \texttt{e5-small-v2}. A query $q$ and a product $d$ are mapped to fixed-length vectors $E_q$ and $E_d$ in the same space, and relevance is measured by their cosine similarity $\cos(E_q, E_d)$.

This design keeps retrieval fast. All product embeddings are computed offline and stored in an ANN index. At query time, only the query is encoded, and its nearest neighbors are looked up among the product vectors. Online latency therefore stays low even over millions of products (Section~\ref{sec:implementationConsiderations}).

\subsection{Input Representation}
Each product is represented by concatenating its main attributes: \texttt{title}, \texttt{brand}, \texttt{item\_type}, \texttt{category}, and an optional \texttt{highlights} field with descriptive text. Each field is preceded by its own marker token, taken from unused tokens in the model vocabulary. The marker embeddings are learned during training and help the encoder recognize field boundaries. For example, a bath towel is encoded as \texttt{[title]} \textit{Performance Plus Oversized Bath Towel} \texttt{[brand]} \textit{Threshold} \texttt{[item\_type]} \textit{Bath Towels} \texttt{[category]} \textit{Bath} \ldots\ Queries are short free-form text and are encoded without field markers.

We also apply \emph{dynamic field masking}. During training, one or more product fields, such as \texttt{brand} or \texttt{category}, are randomly dropped from each example. This keeps the model from depending too heavily on any single attribute and teaches it to match products whose metadata is incomplete or noisy.

\subsection{Loss Function}
We train the dual encoder with an \textit{improved contrastive loss} (ICL) that adds more negative relationships to the standard contrastive objective. In InfoNCE \cite{oord2018representation}, each query $q_i$ in a batch is paired with its positive product $d_i$ and trained to score this pair above the pairs $(q_i, d_j)$ for $j \neq i$.

Following Li et al.~\cite{li2023generaltextembeddingsmultistage}, we use a bidirectional version of this objective. For a batch of $n$ query-product pairs $\{(q_i, d_i)\}_{i=1}^n$, the loss is
\begin{equation}
\label{equ:icl}
    \mathcal{L}_{\mathrm{ICL}} = -\frac{1}{n} \sum_{i=1}^n \log \frac{e^{s(q_i, d_i)/\tau}}{Z_i}
\end{equation}
where $s(\cdot,\cdot)$ is cosine similarity and $\tau$ is a temperature. The partition term $Z_i$ includes all cross-similarities in the batch:
\begin{equation}
\begin{aligned}
    Z_i = {} & \sum_{j} e^{s(q_i,d_j)/\tau} + \sum_{j\neq i} e^{s(q_i,q_j)/\tau} \\
    & + \sum_{j} e^{s(q_j,d_i)/\tau} + \sum_{j\neq i} e^{s(d_j,d_i)/\tau}
\end{aligned}
\end{equation}

Besides the positive pair $(q_i,d_i)$, the partition term treats every $(q_i,d_j)$ and $(q_j,d_i)$ with $j \neq i$ as negatives, along with query-query and product-product similarities within the batch. The first two sums use the query as the anchor, and the last two use the product \cite{li2023generaltextembeddingsmultistage}. As a result, the loss also pushes apart unrelated queries and unrelated products, in addition to separating each query from the wrong products.

A triplet loss \cite{schroff2015facenet} uses one positive and one or a few negatives per example and leaves most of the batch unused. A softmax over the full catalog, with each product as a class, would be impractical at our catalog size. ICL uses every example in the batch as a negative on both the query side and the product side. Li et al. report that this objective produces better embeddings than InfoNCE and one-directional variants, and Table~\ref{tab:model-ablation} shows the same pattern on our data.

\subsection{Training Strategy}
We fine-tune in two stages of increasing difficulty. In the first stage, we train with random negatives to give the model a coarse sense of relevance. Each training instance contains a query $q$, a positive product $d^+$, and $n$ negatives $d^-$ drawn uniformly from the catalog or from other queries' positives. These negatives are clearly irrelevant, and several epochs on them establish an initial embedding space.

In the second stage, we continue training with the hard negatives described in Section~\ref{sec:data}, following prior work showing that mined hard negatives substantially improve dense retrieval \cite{xiong2021approximate}. Exposure-based negatives are products that users saw repeatedly and ignored, and similarity-based negatives are near misses that share surface features with the query. Training on them teaches the model to reject products that look relevant but are not, which improves precision at the top of the ranking.

\section{Precision Control}
\label{sec:precisionControl}

Vector search can return items that are close to the query in embedding space but wrong on an attribute the user cares about. To reduce these off-target results, we add a \textit{precision control module} that filters vector candidates early in the retrieval pipeline.

The module combines a query classifier with a Named Entity Recognition (NER) model to extract both explicit and implicit intent from the query, and then applies high-confidence filters to the candidate set. We define five filters, one for each of the following item dimensions: item type, category, brand, gender, and size group.

The NER model captures \textit{explicit intent} stated in the query, such as gender, brand, or size. The query classifier infers \textit{implicit intent}, mainly item type and category, which a query often implies without naming.

We also use precision control to keep \textit{inappropriate items} out of the results. These are products that are close to the query embedding but wrong for the context or unsuitable for a general audience. For example, a search for food should not return pet products unless the query asks for them. We handle this with a \textit{tiered filtering mechanism} based on item type and category that limits exposure of sensitive or ambiguous categories. Sensitive items appear only when a user searches for them explicitly, which reduces mismatches and helps maintain trust in search results.

The filters are served by a standalone service behind a REST API. At serving time, this API is called in real time to produce filter parameters, which are applied to the candidate pool before final ranking. Keeping the filters in a separate service makes them easy to integrate and tune. Before deployment, we tuned the threshold for each filter to trade off precision against recall. Section~\ref{sec:experimentsAndResults} evaluates the effect of the filters on retrieval quality.

\section{Merging Strategy}
\label{sec:mergingStrategy}

We compared two ways to merge the results of the lexical and vector channels: \textbf{Reciprocal Rank Fusion (RRF)} \cite{Cormack2009} and \textbf{Weighted Interleaving}.

\subsection{Reciprocal Rank Fusion (RRF)}
RRF assigns each candidate document $d$ the score
\[
s(d) \;=\; \sum_{k} \frac{1}{r_{k}(d) + C},
\]
where $r_{k}(d)$ is the rank of $d$ in channel $k$ and $C$ is a smoothing constant, set to 60 in our experiments. RRF rewards agreement between lists. A document that appears in both lists usually scores higher than one that appears in only one list, even if the latter is ranked slightly higher there. The method works best when the input lists overlap substantially, so that the ranks of shared items reinforce each other.

\subsection{Weighted Interleaving}
Weighted interleaving gives each channel a global weight and builds the merged list one position at a time. At each position, a channel is drawn at random according to the weights, and the highest-ranked remaining item from that channel is added to the list. For two channels $A$ and $B$ with weights $w_A + w_B = 1$, the next item comes from $A$ with probability $w_A$ and from $B$ with probability $w_B$. Items already placed from the other channel are skipped, so each product appears at most once. The process repeats until the list reaches the required length.

This method assumes that each list is well ranked on its own. It suits cases where the lists overlap little, because it keeps each channel's ordering intact and gives each channel a share of the results in proportion to its weight.

In our setting the two channels overlap little, and in online A/B tests weighted interleaving outperformed RRF (Section~\ref{sec:experimentsAndResults}). We therefore use it in production. We tuned the channel weights through a series of A/B tests in an explore-then-commit fashion. Starting from equal weights for the lexical and vector channels, we adjusted the ratio across successive experiments and evaluated each setting on the engagement and business metrics described in Section~\ref{sec:experimentsAndResults}. A setting was committed to production once its A/B test showed a statistically significant positive lift ($p < 0.05$) on the key metrics. The weight values are withheld for confidentiality.

\section{Implementation Considerations}
\label{sec:implementationConsiderations}

Running the semantic retrieval model inside a production search stack raised several practical questions: how to serve the model, how to scale it, how to keep inference fast, what hardware to use, and how to support both real-time and batch processing. This section covers each of them.

\subsection{Embedding Service Design}
Query and product embeddings are served in real time by a dedicated \textit{embedding service} that hosts our fine-tuned dual encoder, which is based on \texttt{e5-small-v2} and trained on the engagement-derived labels described in Section~\ref{sec:data}. We chose the model by weighing retrieval quality against model complexity, meaning inference latency and throughput. In isolated CPU benchmarks, \texttt{e5-small-v2} (about 33M parameters, 12 layers, 384-dimensional output) averaged about 25\,ms per query, while the larger \texttt{e5-base-v2} and \texttt{e5-large-v2} had higher latency for a small gain in NDCG (Table~\ref{tab:pretrained-models}). The 6-layer \texttt{all-MiniLM-L6-v2} gave up to 3$\times$ higher throughput but scored lower on our benchmark.

Latency measured in isolation differs from latency under production load, where concurrency, batching, and autoscaling all come into play. Our serving requirement was a p99 latency below 50\,ms at all times. \texttt{e5-small-v2} meets this requirement in production, with p75 and p95 latencies of roughly 25 to 30\,ms. Its 384-dimensional embeddings also keep memory use in the vector index low and reduce serialization cost during retrieval. Because the native embedding size already met our latency and throughput requirements on CPU, we did not explore further dimensionality reduction such as PCA or Matryoshka representations.

\begin{figure}[t]
\centering
\includegraphics[width=0.47\textwidth]{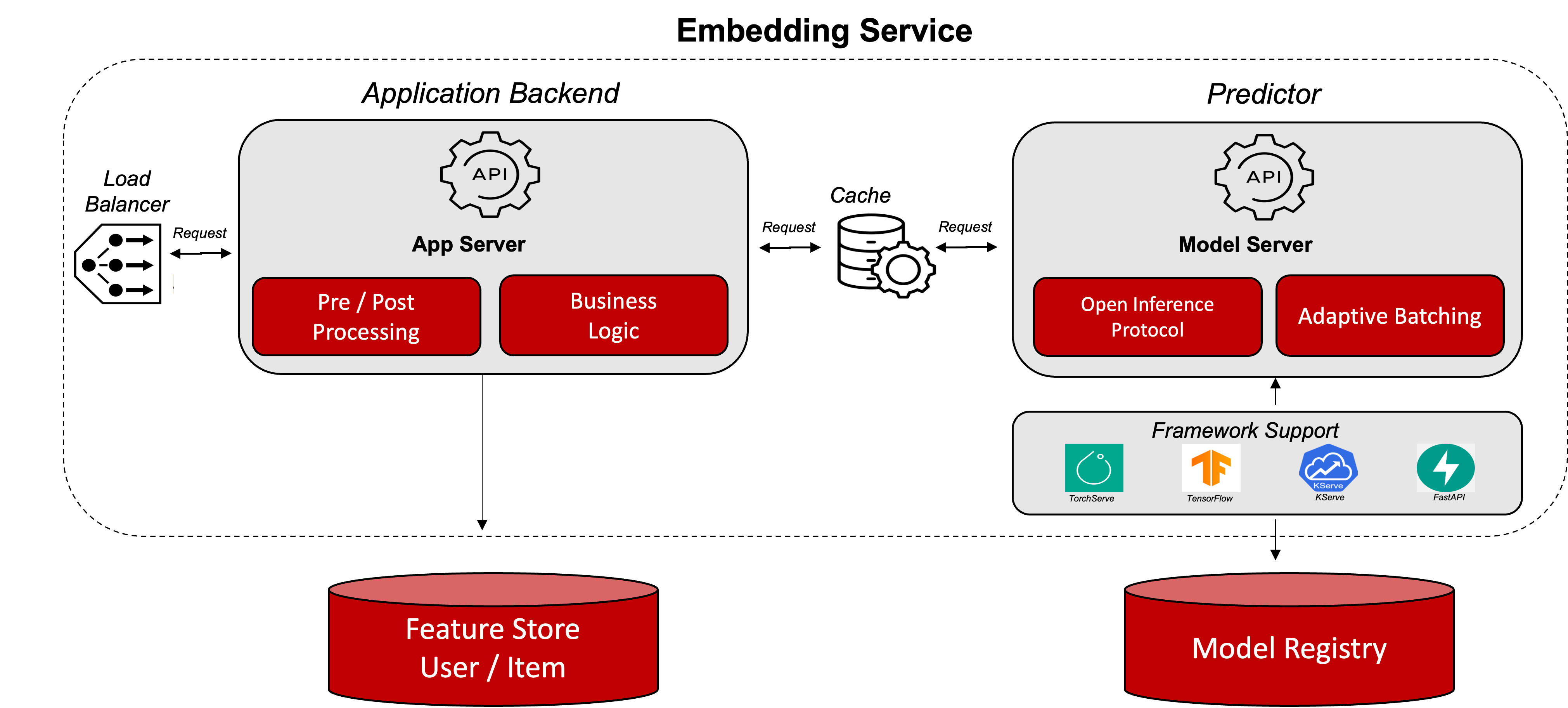}
\caption{Architecture of the real-time embedding service. A front-end application server (left) receives queries from the load balancer, performs preprocessing, and manages business logic. It communicates with a back-end model server (right) running the embedding model (via TorchServe). A caching layer intermediates between the two to store and retrieve frequently used embeddings. The model server supports standard inference protocols and can adaptively batch requests to optimize throughput.}
\Description{System architecture diagram showing a front-end application server connected to a back-end embedding model server through a caching layer. The front-end receives queries from a load balancer, performs preprocessing and business logic, and sends requests for embeddings. The back-end runs the model through TorchServe and supports inference and adaptive batching. The cache stores and retrieves frequently used embeddings between the two servers.}
\label{fig:emb-service-arch}
\end{figure}

The embedding service is a standalone microservice (Figure~\ref{fig:emb-service-arch}). A lightweight application layer handles request routing, input preprocessing, post-processing, and business logic such as feature augmentation and A/B testing hooks. It calls a model server built on TorchServe, which hosts the fine-tuned model (\texttt{e5-finetuned-v2}) as a PyTorch/TorchScript artifact behind a REST API. Separating the online API from the model runtime lets us swap or upgrade models with few changes to query processing.

To avoid repeated computation, the service keeps an in-memory cache of embeddings for frequent and recent queries, so repeated searches skip model inference. This helps most for head queries that recur across many users. The model server also uses adaptive batching, grouping concurrent requests into a single forward pass to reduce compute cost under heavy load.

In production, the embedding service runs on CPU-based infrastructure. Search traffic on our platform reaches millions of requests per day, and load tests confirmed that the service handles peak traffic within its latency budget. Model hosting, caching, and micro-batching together let the service meet these targets at reasonable cost and leave room for future model updates.

\subsection{Scalability}
The semantic retrieval pipeline has to handle high query volume and a growing index. Our platform serves millions of search queries per day, and peak hours reach several times the average load. To absorb these peaks without slowing responses, the model serving layer runs on Kubernetes. The service is containerized and replicated across a cluster, so it scales horizontally with query volume. An autoscaling policy monitors query rate and latency and adds pods when traffic rises, so that no single instance is overloaded.

The vector index also has to scale with the catalog. It holds embeddings for millions of products, and new products and content changes arrive throughout the day. The ANN index is distributed across multiple shards and uses ScaNN for similarity search (Section~\ref{sec:architecture}). Updates are applied in near real time through an asynchronous pipeline, so new product embeddings from the indexing pipeline reach the ANN index with little delay and search results stay fresh as the catalog changes.

\subsection{Infrastructure}
We chose hardware to balance cost and performance for training and serving. Fine-tuning \texttt{e5-small-v2} on about 20 million query-product pairs ran on NVIDIA A100 GPUs with 80\,GB of memory, with distributed training across multiple instances. The large GPU memory allowed large batch sizes and longer input sequences, so the model could use more of each product description. Serving, both online and batch, runs on CPU instances without GPUs. This works because \texttt{e5-small-v2} is small enough to run efficiently on modern multi-core CPUs, especially when exported to TorchScript, and CPU instances are cheaper to scale to many replicas. GPUs are used where they help most, in offline training, and commodity CPU clusters handle serving, a split that is common in industrial search systems.

\section{Experiments and Results}
\label{sec:experimentsAndResults}

We evaluate the system offline and in online A/B tests. This section describes the evaluation data and metrics, ablations on training data and modeling choices, the precision control experiments, and the online results. Test queries do not overlap with training or validation queries, so the offline evaluation measures performance on queries the model has not seen.

\subsection{Offline Evaluation}

In production the model retrieves from the full catalog. Indexing the full catalog for every model variant was not practical, so offline we evaluate each model on how well it ranks a fixed candidate pool for each query. The pools come from an engagement-labeled golden set of more than 5,000 queries built from transaction logs. Each (query, item) pair is labeled on an ordinal scale from 0 to 4 based on observed behavior such as clicks, add-to-cart actions, and purchases, with higher grades for stronger engagement. Each query has about 200 candidate products, both relevant and irrelevant, and the model must rank them. This set is separate from the human-labeled benchmark used for model selection in Section~\ref{sec:semanticModel}.

We report NDCG at several cutoffs and Mean Reciprocal Rank (MRR) \cite{jarvelin2002cumulated,Manning2008}. NDCG@K measures how well the model places highly graded items near the top, and MRR measures how early the first relevant item appears. Both matter in e-commerce, where users often engage only with the first few results. For the precision control experiments we report Precision@K, the fraction of the top $K$ results that are desirable, which directly measures whether the filters remove off-target items.

This setup measures ranking quality within a candidate pool, not recall from the full catalog. We assume that models that rank the golden set well will also retrieve well from the full catalog. The online results in Section~\ref{sec:onlineEval}, where offline gains carried over to live traffic, are consistent with this assumption.

\subsection{Data Processing Experiments}

We study how training data volume, stratified sampling, and hard negatives affect retrieval quality. All rows in Table~\ref{tab:data-ablation} use the improved contrastive loss and the four base product fields (title, brand, item type, and category), so only the training data changes from row to row.

Expanding the training set from 7M to 20M examples gave a large improvement, with NDCG@5 rising from 0.8200 to 0.8661 and MRR from 0.9473 to 0.9760. Broader coverage of queries and products helps the model generalize to unseen queries.

Stratified sampling on its own changed the offline metrics by less than 0.001 in either direction. We kept it because it balances category and query-frequency coverage in the training data (Section~\ref{sec:data}), which the golden set does not measure directly.

Adding hard negatives gave the best result on every metric. NDCG@1 rose to 0.8613, NDCG@5 to 0.8794, and MRR to 0.9822, which shows the value of informative negatives in contrastive training.

\begin{table}[t]
\centering
\caption{Ablation over training data. Each row adds one change to the data pipeline. All rows use the improved contrastive loss and the four base product fields. Bold marks the best value in each column.}
\label{tab:data-ablation}
\setlength{\tabcolsep}{2pt}
\small
\begin{tabular}{>{\raggedright\arraybackslash}p{0.20\linewidth}
                >{\raggedleft\arraybackslash}p{0.14\linewidth}
                >{\raggedleft\arraybackslash}p{0.14\linewidth}
                >{\raggedleft\arraybackslash}p{0.14\linewidth}
                >{\raggedleft\arraybackslash}p{0.14\linewidth}
                >{\raggedleft\arraybackslash}p{0.14\linewidth}}
\toprule
\textbf{Data Variant} & \textbf{NDCG @1} & \textbf{NDCG @5} & \textbf{NDCG @8} & \textbf{NDCG @24} & \textbf{MRR} \\
\midrule
7M Data & 0.8005 & 0.8200 & 0.8290 & 0.8720 & 0.9473 \\
20M Data & 0.8465 & 0.8661 & 0.8744 & 0.9071 & 0.9760 \\
20M Data + \\ Stratified Sampling & 0.8459 & 0.8664 & 0.8752 & 0.9077 & 0.9761 \\
20M Data + \\ Stratified Sampling + \\ Hard Negatives & \textbf{0.8613} & \textbf{0.8794} & \textbf{0.8871} & \textbf{0.9163} & \textbf{0.9822} \\
\bottomrule
\end{tabular}
\end{table}

\subsection{Modeling Experiments}

Table~\ref{tab:model-ablation} isolates modeling choices. We first compare loss functions after one epoch of training. Replacing InfoNCE with the improved contrastive loss improves every metric, with NDCG@5 rising from 0.7621 to 0.7957 and MRR from 0.9201 to 0.9469. The improved loss learns better query and product embeddings even with limited training.

Rows 3 and 4 use the improved contrastive loss with full training on the complete training set. Row 3 adds field marker tokens to the four base fields. Because row 3 also moves from one epoch to full training, the gain from row 2 to row 3 reflects both changes. With field tokens, NDCG@1 reaches 0.8528, NDCG@5 0.8718, and MRR 0.9786.

Row 4 adds the \texttt{highlights} field and dynamic field masking, in which whole fields, including highlights, are randomly dropped during training (Section~\ref{sec:semanticModel}). Highlights add descriptive detail that the base fields lack, and masking keeps the model from relying too heavily on this long field. This configuration performs best on every metric, with NDCG@1 of 0.8622, NDCG@5 of 0.8803, and MRR of 0.9819, and it is the model deployed as Finetuned~v2. Tables~\ref{tab:data-ablation} and~\ref{tab:model-ablation} come from separate ablation runs, so values should be compared within each table rather than across the two.

\begin{table}[t]
\centering
\caption{Ablation over the training objective and product representation. Rows 1 and 2 compare losses after one epoch. Rows 3 and 4 use full training and add changes cumulatively. Row 4 is the configuration deployed as Finetuned~v2. Bold marks the best value in each column.}
\label{tab:model-ablation}
\setlength{\tabcolsep}{2pt}
\small
\begin{tabular}{>{\raggedright\arraybackslash}p{0.24\linewidth}
                >{\raggedleft\arraybackslash}p{0.13\linewidth}
                >{\raggedleft\arraybackslash}p{0.13\linewidth}
                >{\raggedleft\arraybackslash}p{0.13\linewidth}
                >{\raggedleft\arraybackslash}p{0.13\linewidth}
                >{\raggedleft\arraybackslash}p{0.13\linewidth}}
\toprule
\textbf{Model Variant} & \textbf{NDCG @1} & \textbf{NDCG @5} & \textbf{NDCG @8} & \textbf{NDCG @24} & \textbf{MRR} \\
\midrule
InfoNCE Loss \\ (1 epoch) & 0.7503 & 0.7621 & 0.7815 & 0.8628 & 0.9201 \\
Improved Contrastive Loss \\ (1 epoch) & 0.7870 & 0.7957 & 0.8123 & 0.8825 & 0.9469 \\
Improved Contrastive Loss \\ + Field Tokens & 0.8528 & 0.8718 & 0.8799 & 0.9113 & 0.9786 \\
Improved Contrastive Loss \\ + Field Tokens \\ + Highlights Field \\ + Dynamic Field Masking & \textbf{0.8622} & \textbf{0.8803} & \textbf{0.8877} & \textbf{0.9166} & \textbf{0.9819} \\
\bottomrule
\end{tabular}
\end{table}

\subsection{Precision Control Experiments}

We compared three filter settings: no filters; NER filters for gender, size, and brand; and NER filters plus behavior filters, which use query classifier predictions for item type and category. Table~\ref{tab:filter-precision} shows that each added set of filters improves precision at every cutoff. P@1 rises from 0.890 with no filters to 0.914 with NER filters and 0.929 with both, and P@10 rises from 0.836 to 0.857 and 0.885. The attribute filters remove mismatched items that the vector channel would otherwise return. Production uses the full filter set.

\begin{table}[!ht]
\centering
\caption{Effect of precision control filters on the precision of vector retrieval. NER filters enforce attributes stated in the query (gender, size, brand). Behavior filters add constraints predicted by the query classifier (item type, category). Bold marks the best value in each column.}
\label{tab:filter-precision}
\setlength{\tabcolsep}{2pt}
\small
\begin{tabular}{>{\raggedright\arraybackslash}p{0.30\linewidth}
                >{\raggedleft\arraybackslash}p{0.12\linewidth}
                >{\raggedleft\arraybackslash}p{0.12\linewidth}
                >{\raggedleft\arraybackslash}p{0.12\linewidth}
                >{\raggedleft\arraybackslash}p{0.12\linewidth}
                >{\raggedleft\arraybackslash}p{0.12\linewidth}}
\toprule
\textbf{Filter Setting} & \textbf{P@1} & \textbf{P@3} & \textbf{P@5} & \textbf{P@8} & \textbf{P@10} \\
\midrule
No Filters & 0.890 & 0.878 & 0.863 & 0.847 & 0.836 \\
NER Filters & 0.914 & 0.896 & 0.882 & 0.866 & 0.857 \\
NER + Behavior Filters & \textbf{0.929} & \textbf{0.920} & \textbf{0.904} & \textbf{0.890} & \textbf{0.885} \\
\bottomrule
\end{tabular}
\end{table}

\subsection{Online Evaluation}
\label{sec:onlineEval}

To measure business impact, we ran A/B tests on live traffic that compared hybrid search, using the fine-tuned vector model, against the existing lexical search. Each experiment ran for 4 weeks, and all reported results are statistically significant at $p < 0.05$. We also used these tests to choose the merging strategy. We tracked the following metrics:

\begin{itemize}
    \item \textbf{Click-through rate (CTR):} initial engagement with search results.
    \item \textbf{Add-to-cart (ATC) rate:} purchase intent and downstream engagement.
    \item \textbf{Order conversion rate:} share of search sessions that end in a purchase.
    \item \textbf{Demand per visitor:} total order value divided by the number of visitors.
    \item \textbf{Relevancy:} how well the returned results match the query intent, measured online through rank-weighted click through rate (CTR).
\end{itemize}

We track engagement and commercial metrics side by side because they can disagree. A model can raise CTR while lowering conversion, for example, which would mean it surfaces products that attract clicks but are not what users want to buy.

As Table~\ref{tab:online-eval} shows, the pretrained \texttt{e5} model deployed zero-shot, without domain fine-tuning, was negative on every metric relative to the lexical baseline, with CTR, ATC, and order conversion down between 4\% and 7\%. The zero-shot model is trained for general semantic similarity rather than for the products users choose to buy. Fine-tuning on engagement-derived labels closes this gap, as the remaining rows show.

\begin{table}[t]
\centering
\caption{Online A/B test results as relative lift (\%) over the lexical search baseline. Each experiment ran for 4 weeks on live traffic. All results are statistically significant at $p < 0.05$. Bold marks the deployed variant.}
\label{tab:online-eval}
\setlength{\tabcolsep}{2pt}
\small
\begin{tabular}{>{\raggedright\arraybackslash}p{0.24\linewidth}
                >{\raggedleft\arraybackslash}p{0.13\linewidth}
                >{\raggedleft\arraybackslash}p{0.13\linewidth}
                >{\raggedleft\arraybackslash}p{0.13\linewidth}
                >{\raggedleft\arraybackslash}p{0.13\linewidth}
                >{\raggedleft\arraybackslash}p{0.13\linewidth}}
\toprule
\textbf{Model Variant} & \textbf{CTR (\%)} & \textbf{ATC/ Visitor (\%)} & \textbf{Order Conv. (\%)} & \textbf{Demand/ Visitor (\%)} & \textbf{Relevancy (\%)} \\
\midrule
Pretrained & $-4.20$ & $-6.20$ & $-6.57$ & $-7.14$ & $-3.81$ \\
Finetuned v1 (RRF) & $-0.85$ & $+1.25$ & $-1.20$ & $+0.49$ & $-0.67$ \\
Finetuned v1 (Interleaving) & $+0.40$ & $+0.44$ & $+0.74$ & $+1.11$ & $+0.13$ \\
Finetuned v2 (Interleaving) & $\mathbf{+0.97}$ & $\mathbf{+0.89}$ & $\mathbf{+0.98}$ & $\mathbf{+1.10}$ & $\mathbf{+0.21}$ \\
\bottomrule
\end{tabular}
\end{table}

To choose the merging strategy, we deployed our first fine-tuned model, Finetuned~v1, and tested it with RRF and with weighted interleaving. Finetuned~v1 was trained on in-house engagement data without explicit hard negatives, and it represented products with the four base fields only, without field tokens or the highlights field. With RRF the results were mixed. ATC per visitor rose by 1.25\%, but CTR and order conversion fell by 0.85\% and 1.20\%. With weighted interleaving all five metrics improved, including CTR ($+0.40$\%) and order conversion ($+0.74$\%). We therefore adopted weighted interleaving in production.

Finetuned~v2 combines the changes evaluated offline: expanded training data with hard negatives, field marker tokens, and the highlights field with dynamic masking, merged with weighted interleaving. It gave the largest gains in CTR ($+0.97$\%), order conversion ($+0.98$\%), and Relevancy ($+0.21$\%), with ATC per visitor up 0.89\%. Demand per visitor rose 1.10\%, on par with Finetuned~v1 with interleaving ($+1.11$\%).

Table~\ref{tab:hybrid-search-examples} shows example queries that are hard for exact term matching. For \textit{``spice lays chips''} and \textit{``big towels bath''}, the system returns products described with different words (\textit{jalapeño}, \textit{oversized}), which shows that it handles \textbf{synonyms}. It also handles \textbf{natural-language queries} such as \textit{``something to keep my feet warm during winter''} and \textit{``easy meals to make in an air fryer.''} For the misspelled query \textit{``yoge mat''}, it still returns a yoga mat, so the dense channel tolerates \textbf{typos} that exact term matching alone would miss.

\begin{table}[t]
\centering
\caption{Top retrieved item for example queries covering synonyms (\textit{spice}, \textit{big towels}), natural-language queries (\textit{keep my feet warm}, \textit{easy meals}), and a misspelling (\textit{yoge}).}
\label{tab:hybrid-search-examples}
\setlength{\tabcolsep}{2pt}
\small
\begin{tabular}{>{\raggedright\arraybackslash}p{0.35\linewidth}
                >{\raggedright\arraybackslash}p{0.60\linewidth}}
\toprule
\textbf{Query} & \textbf{Top Retrieved Item} \\
\midrule
spice lays chips & Lay's Kettle Cooked Jalapeño Flavored Potato Chips \\
\addlinespace
big towels bath & Performance Plus Oversized Bath Towel Tan – Threshold™: 100\% Cotton \\
\addlinespace
something to keep my feet warm during winter & Costway Foot \& Calf Massager – Deep Kneading Shiatsu Massager Machine with Heat \\
\addlinespace
easy meals to make in an air fryer & Frozen Chicken and Vegetable Potstickers – 12oz – Good \& Gather™ \\
\addlinespace
yoge mat & Yoga Mat 3mm Teal – All In Motion™: Non-Slip PVC, Solid Pattern \\
\bottomrule
\end{tabular}
\end{table}

\subsection{Zero-Result Rate Reduction}

We also tracked the \emph{zero-result rate}, the fraction of searches that return no products. A zero-result search is a complete retrieval failure and is strongly associated with users abandoning the search. In the Finetuned~v2 A/B test, we measured this rate in both the lexical-only control and the hybrid treatment over the full 4 weeks. Hybrid search reduced zero-result searches by about 50\% relative to the control. The vector channel can return related products for queries that have no lexical match in the catalog, which is common for natural-language, misspelled, and long-tail queries. Some zero-result searches remain, for example when precision control removes every vector candidate because none matches the attributes in the query.

The online results are consistent with the offline findings. The gains from fine-tuning the dual encoder on engagement-derived labels with structured item text carried over to live traffic, and Finetuned~v2 improved every tracked metric over the lexical-only baseline, including a $+1.10\%$ lift in demand per visitor. Following company policy, we report relative lifts only.

\section{Conclusion}
\label{sec:conclusion}

This paper presented a hybrid search system at Target that combines lexical and semantic retrieval for product search. The system is deployed on Target's web and mobile platforms and serves millions of search queries each day. Against the lexical-only baseline, the best variant in our A/B tests raised CTR by 0.97\%, order conversion by 0.98\%, and demand per visitor by 1.10\%, and it cut zero-result searches by about half (Section~\ref{sec:experimentsAndResults}).

Lexical and dense retrieval complemented each other in our setting. Lexical retrieval remains strong when the query shares terms with the product text, and the dense channel finds relevant products when it does not, as the qualitative examples and the drop in zero-result searches show. Because the two channels return largely different items, weighted interleaving merged them better than RRF, which depends on overlap between lists.

Domain adaptation mattered more than model size. Moving from \texttt{e5-small-v2} to \texttt{e5-large-v2} added less than 0.01 NDCG on our benchmark (Table~\ref{tab:pretrained-models}), while the zero-shot model lost engagement online and the fine-tuned models gained (Table~\ref{tab:online-eval}). Offline, training data volume, hard negatives, and structured product input each added measurable gains (Tables~\ref{tab:data-ablation} and~\ref{tab:model-ablation}).

Serving constraints shaped the design from the start. Caching, adaptive batching, and ANN indexing kept the embedding service within its latency budget on CPUs, and the precision control module kept vector results consistent with the attributes stated in or implied by the query (Table~\ref{tab:filter-precision}).

\section{Future Work}
\label{sec:futureWork}

We see three directions that build on this architecture and address limitations we observed in production.

\subsection{Personalized Retrieval}
The current system returns the same candidates to every user who issues the same query. One direction is to bring user-level signals into retrieval, for example by conditioning the query embedding on a user's browsing and purchase history so that the vector channel reflects individual preferences. A user whose past purchases lean toward organic products could then see more organic options for a generic grocery query. The main design question is how to balance personalization with diversity. Narrow personalization can limit product discovery, especially for new or infrequent shoppers, so a personalized retrieval score would likely need to be paired with a diversity-aware re-ranking objective.

\subsection{Multimodal and Multi-Channel Retrieval}
Text alone does not capture everything that makes a product relevant. Product images carry information about style, color, and shape that is hard to express in textual attributes. Adding image embeddings alongside text would allow cross-modal retrieval, in which a text query can surface visually matching products and an image query can retrieve products described in text. This requires a shared embedding space for text and images, and vision-language models fine-tuned on catalog data are a natural candidate. Further retrieval channels, such as behavioral co-occurrence and structured attribute matching, could also be added to the existing merging framework to broaden recall.

\subsection{Learned Precision Control}
The precision control module in Section~\ref{sec:precisionControl} relies on a fixed set of NER-based and classifier-based filters. Their thresholds are tuned offline and do not adapt to the query at inference time. A lightweight learned model could instead score each vector candidate's relevance to the query, replacing hard thresholds with soft scores. This could handle ambiguous or multi-intent queries better and reduce both false positives, where irrelevant items pass the filter, and false negatives, where relevant items are removed.

\begin{acks}
We thank the Search Data Science leadership at Target for supporting this work. We are grateful to the Search Product team for helping formulate the hybrid search use case, design the A/B tests, and analyze the results. We also thank the Search Ranking and Engineering teams for their work on the ranking service and the vector database that this system builds on.
\end{acks}

\bibliographystyle{ACM-Reference-Format}
\bibliography{References}

\end{document}